\documentclass[10pt,a4paper,twocolumn]{article}

\usepackage{graphicx}
\usepackage{amsmath}
\usepackage{ amssymb }
\usepackage{authblk}
\newcommand{\sna}{\{n_{\alpha}\}}
\newcommand{\sa}{\alpha}
\newcommand{\wa}{\omega_{\alpha}}
\newcommand{\npal}{n_{p_{\alpha}}}
\newcommand{\npap}{n_{p_{\alpha}+1}}
\newcommand{\br}{\textbf{r}}
\newcommand{\vp}{v_\mathrm{p}}

\begin{document}
\title{Exact partition potential for model systems of interacting electrons in 1-D
}
\date{}
\author[1]{Yan Oueis
}
\author[1,2]{Adam Wasserman
}

\affil[1]{Department of Chemistry, Purdue university, West Lafayette, IN 47907 USA
}
\affil[2]{Department of Physics and Astronomy, Purdue university, West Lafayette, IN 47907 USA
}

\maketitle

\begin{abstract}
We find the numerically exact partition potential for 1-D systems of interacting electrons designed to model diatomic molecules. At integer fragment occupations, the kinetic contribution to the partition potential develops sharp features in the internuclear region that nearly cancel corresponding features of exchange-correlation. They occur at locations that coincide with those of well-known features of the underlying molecular Kohn-Sham potential. For non-integer fragment occupations, we demonstrate that the fragment Kohn-Sham gaps determine the kinetic part of the partition potential. Our results highlight the importance of non-additive noninteracting kinetic and exchange-correlation energy approximations in density-embedding methods at large internuclear separations and the importance of non-additive noninteracting kinetic energy approximations at \textit{all} separations.
\end{abstract}

%%%%%%%%%%%
%INDRODUCTION%
%%%%%%%%%%%
\section{Introduction}
\label{sec:intro}
The modern approach to the theory of chemical change is deeply rooted in the formalism of density functional theory (DFT). The foundation was built by Parr, Yang, Ayers, Geerlings and others. \cite{PY89,AP00,GPL03,HG12} It is based on the analysis of the change to the ground state properties of isolated molecular fragments induced by other fragments approaching from infinity. \cite{CW07} This approach made it possible to identify some of the most common pre-DFT reactivity indices with functional derivatives of the ground state molecular quantities. Nevertheless, the theory lacks the finite-distance interactions that play an essential role in the fragment chemical behavior. Noticeably, the formulation of Parr's reactivity indices within the non-integer DFT formalism (PPLB formalism) \cite{PPLB82} leads to conceptually inconsistent results. \cite{CW07}

The Partition Theory (PT) of ref. \cite{CW07} aims at solving these inconsistencies. $\mathrm{PT}$ imagines a fictitious system of noninteracting fragments embedded in a global potential (\textit{i.e.} same for all fragments). The fragments are constrained to have densities that sum to the total molecular density while minimizing the sum of fragment energies (more on this quantity later). The uniqueness of the fragment densities is ensured by the global embedding potential, according to the theorem of ref. \cite{CW06}.

To formally introduce the PT, we partition the external potential $v(\br)$ into fragments labeled by the index $\alpha$:
\begin{equation}
v(\br)=\sum_{\sa}{v_\sa(\br)}.
\label{eq:v}
\end{equation}
$\mathrm{PT}$ is based on the following decomposition of the molecular ground state ($\mathrm{GS}$) energy:
\begin{equation}
E_v[n_{\mathrm{GS}}]=\min_{n_N\rightarrow N}[\min_{\sna\rightarrow n_N}[E_\mathrm{f}[\sna]]+E_\mathrm{p}[n_N]],
\label{eq:Evdec}
\end{equation}
where $E_\mathrm{f}[\sna]$ is the sum of fragment energies and $E_\mathrm{p}[n_N]$ is the partition energy. In eq. \ref{eq:Evdec}, the outer minimization is over all densities that integrate to $N$ electrons. Each of the fragment contributions to $E_\mathrm{f}$ is defined to have the PPLB functional form:
\begin{equation}
\begin{split}
E_\mathrm{f}[\sna]=&\\ &\sum_{\sa}{\{(1-\wa)E_{v_{\sa}}[\npal]+\wa E_{v_{\sa}}[\npap]\}},
\end{split}
\label{eq:Ef}
\end{equation}
where $p_\sa$ and $\wa$ are the integer and fractional parts of $N_\sa$ (number of electrons in fragment $\sa$). The inner minimization in \ref{eq:Evdec} is over all $p_\sa$, $\wa$, $\npal(\br)$, and $\npap(\br)$ that produce the density $n_\mathrm{f}(\br)=n_N(\br)$ according to:
\begin{equation}
n_\mathrm{f}(\br)=\sum_{\sa}{\{(1-\wa)\npal(\br)+\wa\npap(\br)\}}.
\label{eq:n}
\end{equation}
To avoid finite-difference derivatives, it is common to fix the integer part of the occupation numbers and use $\sna$ to denote the set of all $\wa$'s, $\npal(\br)$'s, and $\npap(\br)$'s. We also follow this convention in this text. Therefore, all our derivatives with respect to $\wa$, $\npal(\br)$, or $\npap(\br)$ are not the ``formal" derivatives but rather constrained derivatives that keep the integer part of the corresponding fragment $\sa$ constant.

The inner minimization in eq. \ref{eq:Evdec} is done by the method of Lagrange multipliers. The equivalent unconstrained extremization is done for the following functional:
\begin{equation}
%\begin{split}
G[\sna,\vp(\br)] =E_\mathrm{f}[\sna]+\int d\br \vp(\br)[n_\mathrm{f}(\br)-n_{\mathrm{GS}}],
%\end{split}
\label{eq:G}
\end{equation}
where the \textit{partition potential}, $\vp(\br)$, has been introduced as the Lagrange multiplier that forces condition \ref{eq:n} to be satisfied at each point in space. Eq. \ref{eq:G} also brings out the physical meaning of the fragment densities
\begin{equation}
n_\sa(\br)=(1-\wa)\npal(\br)+\wa\npap(\br).
\label{eq:na}
\end{equation}
They are the ensemble ground state densities of $N_\sa$ electrons in the potential $(v_\sa(\br)$ $+$  $\vp(\br))$. The partition potential $\vp(\br)$ is the above-mentioned global embedding potential that guarantees the uniqueness of the $n_\sa$'s. \cite{CW06} Note that $E_{v_{\sa}}[\npal]$ in eq. \ref{eq:Ef} is \textit{not} the correct ground state energy corresponding to $\npal(\br)$, but $E_{v_{\sa}+\vp}[\npal]$ is. 

Stationarity of $G[\sna,\vp(\br)]$ with respect to $\wa$ implies: \cite{CW06}
\begin{equation}
\mu_\sa^{\mathrm{PT}}=\mu_\beta^{\mathrm{PT}},
\label{eq:mueq}
\end{equation}
for any two fragments $\sa$ and $\beta$, where the $\sa$-chemical potential of $\mathrm{PT}$ is defined as
\begin{equation}
\begin{split}
\mu_\sa^{\mathrm{PT}}=& (E_{v_{\sa}}[\npap]+\int{d\br \vp(\br)\npap(\br)})- \\&(E_{v_{\sa}}[\npal]+\int{d\br \vp(\br)\npal(\br)}).
\end{split}
\label{eq:mu}
\end{equation}

Following the standard Kohn-Sham (KS) decomposition of the energy, the partition energy of eq. \ref{eq:Evdec} can be written as:
\begin{equation}
\begin{split}
E_\mathrm{p}[n_N]=&T_\mathrm{s}^{\mathrm{nad}}[\sna]+E_{\mathrm{ext}}^{\mathrm{nad}}[\sna]+E_{\mathrm{H}}^{\mathrm{nad}}[\sna]+ \\ &E_{\mathrm{XC}}^{\mathrm{nad}}[\sna],
\label{eq:Epdec}
\end{split}
\end{equation}
where $T_\mathrm{s}$ is the noninteracting kinetic energy, $E_{\mathrm{ext}}$ is the interaction energy of electrons with the external potential, $E_\mathrm{H}$ is the Hartree energy, and $E_{\mathrm{XC}}$ is the exchange-correlation energy. The superscript ``$\mathrm{nad}$" indicates that each of these functionals is a non-additive contribution defined (for an arbitrary functional $\Pi$) as: $\Pi^{\mathrm{nad}}[\sna] = \Pi[n_N] - \sum_{\sa}\{(1-\wa)\Pi_\sa[\npal]+\wa \Pi_\sa[\npap]\}$.

The relationship between $E_\mathrm{p}[\sna]$ and $\vp(\br)$ was derived by Nafziger and Wasserman: \cite{NW14}
\begin{equation}
\begin{split}
\vp(\br) =& \int d\br \sum_{\sa}\{   \frac{\delta E_\mathrm{p}}{\delta \npal(\mathbf{r^\prime})}\frac{\delta \npal(\mathbf{r^\prime})}{\delta n_{\mathrm{f}}(\br)} + \\ &\frac{\delta E_\mathrm{p}}{\delta \npap(\mathbf{r^\prime})}\frac{\delta \npap(\mathbf{r^\prime})}{\delta n_{\mathrm{f}}(\br)}  \}.
\end{split}
\label{eq:Epvp}
\end{equation}
Substituting \ref{eq:Epdec} into \ref{eq:Epvp} leads to a useful decomposition of $\vp(\br)$ into contributions from kinetic, external, Hartree, and exchange-correlation parts.

Used with approximate density functionals, $\mathrm{PT}$ has been shown to fix delocalization and static correlation errors in bond-stretching. \cite{NW15} It has also been successfully applied to the construction of approximations to non-additive noninteracting kinetic energy functionals. \cite{NJW17,JNW18} The exact properties of $\mathrm{PT}$ were analyzed with numerically solvable model systems of noninteracting electrons \cite{CWB07,CWCB09,ECWB09,EBCW10,TNW12} but the case of interacting electrons has only been studied approximately. 

Here, for the first time, we solve the exact $\mathrm{PT}$ problem for systems of interacting electrons. We use simple 1-D models of hydrogen dimer ($\mathrm{H_2}$), helium hydride cation ($\mathrm{HeH}^{+}$) and lithium hydride ($\mathrm{LiH}$) molecules. In these model cases, two valence electrons interact via a soft-coulomb potential. \cite{E90,LGE00,BSWBW15} These models can be solved numerically exactly. We use these exact results to study the connection between KS and PT formalisms and the effect of electron-electron interaction on the most prominent features of the partition potential and its components.
%%%%%%%%%%%%%%%%%%
%  COMPUTATIONAL  DETAILS  %
%%%%%%%%%%%%%%%%%%
%%
\section{Model system and numerical methods}
\label{sec:CD}
The properties of each fragment as well as the entire system are computed on a fine real grid. Density-to-potential inversion techniques are used to solve the $\mathrm{PT}$ problem (\textit{i.e.} the problem of finding $\vp(\br)$ for a given density and choice of partitioning). A more detailed discussion of the numerical methods is presented below.
\subsection{Model Hamiltonians}
\label{ssec:sys}
Our model of a 1-D dimer has two interacting valence electrons. The soft-coulomb (SC) potential is used to model charge-charge interactions. The electronic Hamiltonian is:
\begin{equation}
\begin{split}
\mathcal{H} =& \sum_{i=1,2}\biggl\{-\frac{1}{2}\nabla^2_{x_i} - \frac{1}{\sqrt{1.0+(x_i-R_{\mathrm{H}})^2}} - \\ &\frac{Z_{\mathrm{X}}}{\sqrt{1.0+(x_i-R_\mathrm{X})^2}}\biggr\}+\frac{\lambda}{\sqrt{1.0+(x_1-x_2)^2}},
\end{split}
\label{eq:H}
\end{equation}
where $x_i$ is the coordinate of the $i$\textsuperscript{th} electron, $R_\mathrm{X}$ is the position of the nucleus $\mathrm{X}$ ($\mathrm{X}$ stands for either $\mathrm{H}$ or $\mathrm{He}$), $Z_{\mathrm{X}}$ is the nuclear charge and $\lambda$ is the parameter that switches the electron-electron interaction on ($\lambda=1$) or off  ($\lambda=0$). We use the softening parameter value of $1.0$ and a simulation box of $25$ $\mathrm{a.u.}$ The case of $\mathrm{LiH}$ is discussed separately in eq. \ref{eq:HLiH}.

With the nuclear-nuclear interaction given by:
\begin{equation}
V_{\mathrm{nn}}=\frac{Z_{\mathrm{X}}}{\sqrt{3.0+(R_{\mathrm{X}} - R_\mathrm{H})^2}},
\label{eq:Vnn}
\end{equation}
the equilibrium bond-length is $R_0=1.6$ $\mathrm{a.u.}$ for $\mathrm{H_2}$ and $R_0=2.1$ $\mathrm{a.u.}$ for $\mathrm{HeH}^{+}$.

The fragment Hamiltonians have the form:
\begin{equation}
\begin{split}
\mathcal{H}_{p_{\sa}+1} =& \sum_{i=1,2}\biggl\{-\frac{1}{2}\nabla^2_{x_i} - \frac{Z_{\mathrm{X}}}{\sqrt{1.0+(x_i-R_{\mathrm{X}})^2} } + \\ &\vp(x_i)\biggr\}+\frac{\lambda}{\sqrt{1.0+(x_1-x_2)^2}}
\end{split}
\label{eq:Hpap1}
\end{equation}
and
\begin{equation}
\mathcal{H}_{p_{\sa}} = -\frac{1}{2}\nabla^2_{x} - \frac{Z_{\mathrm{X}}}{\sqrt{1.0+(x-R_{\mathrm{X}})^2} } + \vp(x).
\label{eq:Hpa}
\end{equation}
\subsection{Decomposition of $\vp(x)$}
\label{ssec:vp}
With the strategy introduced by eqs. \ref{eq:Epdec} and \ref{eq:Epvp}, we rewrite $\vp(x)$ as:
\begin{equation}
\vp(x) = v_{\mathrm{p},\mathrm{kin}}(x) + v_{\mathrm{p},\mathrm{ext}}(x) + v_{\mathrm{p},\mathrm{H}}(x) + v_{\mathrm{p},\mathrm{XC}}(x),
\label{eq:vpdc}
\end{equation}
where the components correspond to those of $E_\mathrm{p}$ in eq. \ref{eq:Epdec}. To calculate each component, we note:
\begin{subequations}
  \begin{align}
    \frac{\delta T_\mathrm{s}^{\mathrm{nad}}[\npal]}{\delta \npal(x)}=&(1-\wa)(v_\mathrm{s}[\npal](x)-v_\mathrm{s}[n_{\mathrm{GS}}](x)), \\
    \frac{\delta E_\sa^{\mathrm{nad}}[\npal]}{\delta \npal(x)}=&(1-\wa)(v(x)-v_\sa(x)), \\
    \frac{\delta E_\mathrm{H}^{\mathrm{nad}}[\npal]}{\delta \npal(x)}=&(1-\wa)\int{dx_1\frac{n_{\mathrm{GS}}(x_1)-\npal(x_1)}{\sqrt{1.0 +(x_1-x)^2}}}, \\
    \frac{\delta E_\mathrm{XC}^{\mathrm{nad}}[\npal]}{\delta \npal(x)}=&(1-\wa)(v_\mathrm{XC}[n_{\mathrm{GS}}](x)-v_\mathrm{XC}[\npal](x)).
  \end{align}
\label{subeq:vpdc1}
\end{subequations}
The equivalent derivatives with respect to the $\npap$ are omitted for brevity. The functional derivatives in eqs. \ref{subeq:vpdc1} can be readily calculated and used further to obtain $v_{\mathrm{p},\mathrm{kin}}(x)$, $v_{\mathrm{p},\mathrm{ext}}(x)$ and $v_{\mathrm{p},\mathrm{H}}(x)$. The remaining $v_{\mathrm{p},\mathrm{XC}}(x)$ is calculated as a difference between the full $\vp(x)$ and the first three components. For the functional derivative ${\delta \npal(x^\prime)}/{\delta n_{f}(x)}$ in eq. \ref{eq:Epvp}, we use the local approximation: \cite{MW13}
\begin{equation}
\frac{\delta \npal(x^\prime)}{\delta n_{\mathrm{GS}}(x)}\approx \mathcal{Q}_{p_\sa}(x,x^\prime) \equiv \frac{\npal(x^\prime)}{n_{\mathrm{GS}}(x)}\delta(x-x^\prime),
\label{eq:qloc}
\end{equation}
resulting in the following equations for the components:
\begin{subequations}
  \begin{align}
    \begin{split}
    \label{subeq:vpdc2a}
    v_{\mathrm{p},\mathrm{kin}}(x)=&\sum_\sa\{ \wa\mathcal{Q}_{p_{\sa+1}}(x,x)v_\mathrm{s}^{(-)}[\npap](x)+\\ 
    &(1-\wa)\mathcal{Q}_{p_\sa}(x,x)v_\mathrm{s}^{(-)}[\npal](x) \}-\\
    &v_\mathrm{s}^{(-)}[n_{\mathrm{GS}}](x),
    \end{split}\\
    v_{\mathrm{p},\mathrm{ext}}(x)=&\sum_\sa\{ (v(x)-v_\sa(x))\frac{n_\sa(x)}{n_{\mathrm{GS}}(x)} \},\\
    \begin{split}
    v_{\mathrm{p},\mathrm{H}}(x)=&\sum_\sa\biggl\{ \wa\mathcal{Q}_{p_{\sa+1}}(x,x)\\
    &\cdot\int{dx_1\frac{n_{\mathrm{GS}}(x_1)-\npap(x_1)}{\sqrt{1.0 +(x_1-x)^2}}}+\\
    &(1-\wa)\mathcal{Q}_{p_\sa}(x,x)\int{dx_1\frac{n_{\mathrm{GS}}(x_1)-\npal(x_1)}{\sqrt{1.0 +(x_1-x)^2}}} \biggr\},
    \end{split}\\
    \begin{split}
    v_{\mathrm{p},\mathrm{XC}}(x)=&v_\mathrm{XC}^{(-)}[n_{\mathrm{GS}}](x)-\\
    &\sum_\sa\{ \wa\mathcal{Q}_{p_\sa+1}(x,x)v_\mathrm{XC}^{(-)}[\npap](x)+\\ 
    &(1-\wa)\mathcal{Q}_{p_\sa}(x,x)v_\mathrm{XC}^{(-)}[\npal](x) \},
\end{split}
   \end{align}
\label{subeq:vpdc2}
\end{subequations}
where the superscript ``$(-)$" indicates that the $x$-independent constant in $v_\mathrm{s}(x)$ at integer electron number is calculated at the limit from below. Since this approximation satisfies the sum rule, $\sum_\sa\{  \mathcal{Q}_{p_\sa} +  \mathcal{Q}_{p_\sa+1} \}=\delta(x-x^\prime)$, the sum of $\vp(x)$ components yields the \textit{exact} $\vp(x)$. \cite{MW13} Although this local approximation was shown to be reliable for various systems \cite{NW14}, it can still affect the individual components. Finally, we note that since $v_\mathrm{s}^{(-)}[\npal](x)=v_\sa(x)+v_{\mathrm{H}}[\npal](x) + v_{\mathrm{XC}}^{(-)}[\npal](x)+\vp(x)$, eqs. \ref{subeq:vpdc2} can be derived simply by construction.
\subsection{Numerical methods}
\label{ssec:numerical}
\paragraph{Exact diagonalization:} Hamiltonians \ref{eq:H}, \ref{eq:Hpap1} and \ref{eq:Hpa} are all diagonalized on a real grid using the sixth order central finite difference method for the $\nabla^2_{x_i}$ operator. \cite{F88} We note that both \ref{eq:H} and \ref{eq:Hpap1} are symmetric under the particle index interchange and all the eigenstates are either symmetric or antisymmetric. Spatially symmetric solutions correspond to the spin zero state while the antisymmetric spatial solutions correspond to triplet spin states. It therefore becomes clear that we simply need to search for the lowest eigenstate of \ref{eq:Hpap1} or \ref{eq:Hpa}. \cite{HFCVMTR11,SSZ96}% We then use a Matlab built-in \textit{eigs} function to solve for the lowest eigen-pair of the resulting Hamiltonian matrices (size $N^2$-by-$N^2$, where $N$ is the number of grid points). 

\paragraph{Density-to-potential inversions:} To obtain the exact $\vp(x)$, we need to perform a numerical inversion. The following outlines the inversion algorithm employed to find $\vp(x)$ for a particular partitioning at a fixed set of fragment occupation numbers:
\begin{enumerate}
\setcounter{enumi}{-1}
\item \label{itm:0} Start with an initial guess for $\vp(x)$.
\item \label{itm:1} Use eqs. \ref{eq:n}, \ref{eq:Hpap1} and \ref{eq:Hpa} to compute the sum of fragment densities in the presence of $\vp(x)$.
\item \label{itm:2} Calculate the difference between the total molecular density and the sum from \ref{itm:1}.
\item \label{itm:3} Based on the value from 2, decide whether the sum of the fragment densities is sufficiently close to the total molecular density. If it is, the optimization is done; otherwise go to \ref{itm:4}.
\item \label{itm:4}Update $\vp(x)$. Go to step \ref{itm:1}.
\end{enumerate}
We note that the algorithm assumes that the total molecular density can be pre-computed. For the convergence criterion in step \ref{itm:3} we use the value of the following functional at step $k$:
\begin{equation}
\theta^{(k)}[n_\mathrm{f}^{(k)}]=\frac{1}{2^2}\int{dx}[ n_\mathrm{f}^{(k)}(x)-n_{\mathrm{GS}}(x) ]^2,
\label{eq:conv}
\end{equation}
where the factor $2$ in the denominator appears because we have two electrons. For the update in step \ref{itm:4}, we utilize the Broyden's method. \cite{B65} After the algorithm is converged, we methodically vary the occupation numbers to eventually scan the entire set and find the minimum. The initial guess of $\vp(x)=0$ in step \ref{itm:0} and the convergence thresholds of $10^{-14}$ in step \ref{itm:3} are sufficient for obtaining accurate energies. To obtain accurate and smooth potentials, we apply the following procedure. After the initial optimization to $\theta^{(k)}\sim10^{-14}$, we compute $v_{\mathrm{p},\mathrm{kin}}^{(k)}(x)$, $v_{\mathrm{p},\mathrm{ext}}^{(k)}(x)$, $v_{\mathrm{p},\mathrm{H}}^{(k)}(x)$ and $v_{\mathrm{p},\mathrm{XC}}^{(k)}(x)$ using eqs. \ref{subeq:vpdc2}. In particular, we use the exact molecular density to compute derivatives of eq. \ref{subeq:vpdc1} and the current $n_\mathrm{f}^{(k)}(x)$ to compute the factors of eq. \ref{eq:qloc}. We then use the computed potetials to find $\bar{v}_{\mathrm{p},\mathrm{kin}}(x)=\vp^{(k)}(x)-v_{\mathrm{p},\mathrm{ext}}^{(k)}(x)-v_{\mathrm{p},\mathrm{H}}^{(k)}(x)-v_{\mathrm{p},\mathrm{XC}}^{(k)}(x)$ and $\bar{v}_{\mathrm{p},\mathrm{XC}}(x)=\vp^{(k)}(x)-v_{\mathrm{p},\mathrm{ext}}^{(k)}(x)-v_{\mathrm{p},\mathrm{H}}^{(k)}(x)-v_{\mathrm{p},\mathrm{kin}}^{(k)}(x)$. Finally, we construct the new guess for $\vp(x)$ by adding $\bar{v}_{\mathrm{p},\mathrm{kin}}(x)$, $\bar{v}_{\mathrm{p},\mathrm{XC}}(x)$, $v_{\mathrm{p},\mathrm{ext}}^{(k)}(x)$ and $v_{\mathrm{p},\mathrm{H}}^{(k)}(x)$. This new guess is run through a single cycle of the algorithm to return the improved results. This procedure does not significantly improve the energy results. However, it markedly improves the density convergence in the low-density regions and produces smooth potentials. Applying this procedure periodically within our algorithm can converge it to machine precision ($\mathrm{max}|n_\mathrm{f}^{(k)}(x)-n_{\mathrm{GS}}(x)|\sim10^{-16}$). However, no appreciable changes in features of the potentials are observed after the threshold of $\theta\sim10^{-14}$.

Since each fragment can only have up to $2$ electrons, the KS potentials can be obtained analytically. The expressions for the inversions are trivial. \cite{JW18}
%%%%%%%%
%  RESULTS %
%%%%%%%%
\section{Illustrative results and discussion}
\label{sec:res}
\begin{figure*}
  \includegraphics[width=1.0\textwidth]{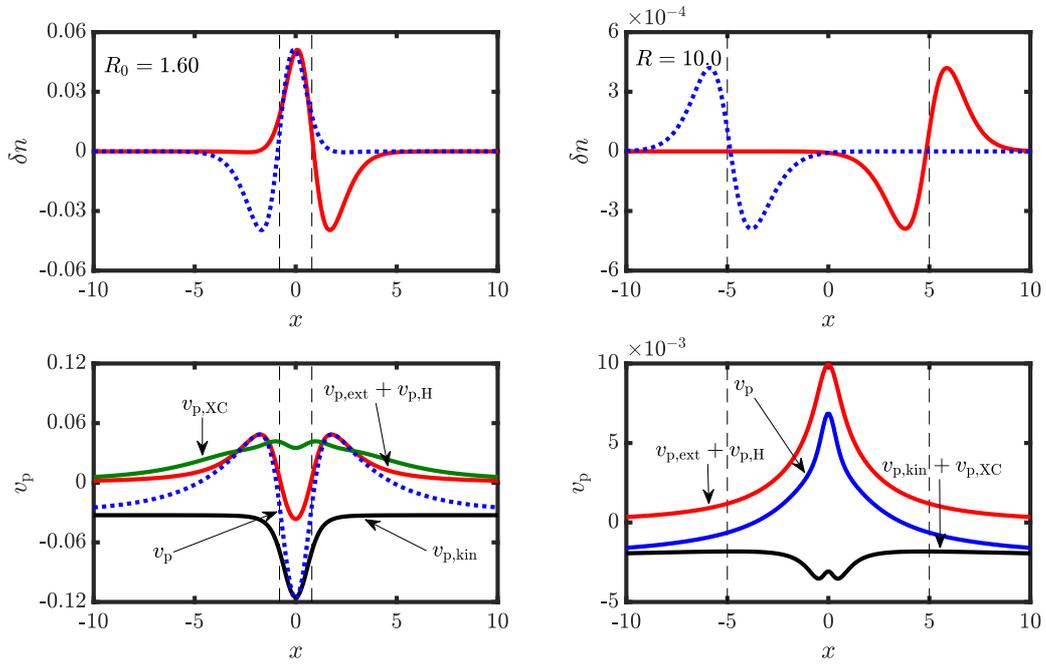}
\caption{1-D $\mathrm{H_2}$ model at $R_0=1.60$ $\mathrm{a.u.}$ (left) and $R=10.0$ $\mathrm{a.u.}$ (right). Top: deformations of the fragment densities $\delta n_\sa(x)=n_\sa(x) - n_\sa^0(x)$, where $n_\sa^0(x)$ is the density of an isolated fragment. Bottom: partition potential $\vp(x)$ and its components defined through eq. \ref{eq:vpdc}. Vertical dashed lines indicate the position of nuclei. The electron-electron interaction parameter $\lambda=1$.}
\label{fig:1} 
\end{figure*}
\paragraph{$\mathbf{H_2}$ model:} We consider first a symmetric dimer model of $\mathrm{H_2}$ at two different internuclear separations: the equilibrium bond length, $R_0=1.60$ $\mathrm{a.u.}$, and the large separation, $R=10.0$ $\mathrm{a.u.}$ The optimal occupations for this model is clearly $N_\mathrm{H,left}=1.0$ and $N_\mathrm{H,right}=1.0$. We analyze features of $\vp(x)$ and how they are affected by the electron-electron interaction. Our results highlight the importance of approximating $v_{\mathrm{p},\mathrm{kin}}(x)$ and $v_{\mathrm{p},\mathrm{XC}}(x)$ accurately in density embedding calculations, as previously pointed out by several computational studies using approximate $T_\mathrm{s}^{\mathrm{nad}}[\npal]$ \cite{NJW17,JNW18,WW93,JN14} For the noninteracting system, we show that $\vp(x)$ is dominated by $v_{\mathrm{p},\mathrm{ext}}(x)$ at $R_0=1.60$ and by $v_{\mathrm{p},\mathrm{kin}}(x)$ $R_0=10.0$.

In fig. \ref{fig:1}, we plot the $\mathrm{PT}$ deformations of the fragment densities ($\delta n_\sa(x)=n_\sa(x) - n_\sa^0(x)$, where $n_\sa^0(x)$ is the density of an isolated fragment) and partition potentials corresponding to these two cases. At $R=10.0$, both densities are slightly shifted away from the interatomic region. In contrast, at the equilibrium separation, the densities are shifted towards the bonding region. Furthermore, the interatomic interactions are markedly weaker at the larger separation. This is reflected in the density deformations and $\vp(x)$ features that are roughly two orders of magnitude smaller than those at the equilibrium bond distance.

In the bottom panels of fig. \ref{fig:1}, we analyze the origin of these features through the decomposition of eq. \ref{eq:vpdc}. We combine $v_{\mathrm{p},\mathrm{ext}}(x)$ and $v_{\mathrm{p},\mathrm{H}}(x)$ because $v_{\mathrm{p},\mathrm{ext}}(x)$ has a deep well and $v_{\mathrm{p},\mathrm{H}}(x)$ has a high peak in the internuclear region. However, their sum is on the order of the features in $\vp(x)$. Adding the external and Hartree components can be further justified by the fact that in practical calculations both can be computed exactly, but $v_{\mathrm{p},\mathrm{kin}}(x)$ and $v_{\mathrm{p},\mathrm{XC}}(x)$ require approximations. In the plot for $R=10.0$, we also combine $v_{\mathrm{p},\mathrm{kin}}(x)$ and $v_{\mathrm{p},\mathrm{XC}}(x)$, as they are analyzed separately later in the paper. At the equilibrium, the depth of the well in $\vp(x)$ is determined by the $v_{\mathrm{p},\mathrm{kin}}(x)$ and the $v_{\mathrm{p},\mathrm{ext}}(x)+v_{\mathrm{p},\mathrm{H}}(x)$ terms. The position of the peaks is also determined by the $v_{\mathrm{p},\mathrm{ext}}(x)+v_{\mathrm{p},\mathrm{H}}(x)$ contribution. We note that the effect of the non-additive XC term is small relative to the other components. At $R=10.0$, the peak in the middle comes from $v_{\mathrm{p},\mathrm{ext}}(x)+v_{\mathrm{p},\mathrm{H}}(x)$. The contribution from $v_{\mathrm{p},\mathrm{kin}}(x)$ is almost completely cancelled by $v_{\mathrm{p},\mathrm{XC}}(x)$, but fine features persist even when the threshold  $\theta^{(k)}$ is decreased to $10^{-23}$.
\begin{figure*}
\includegraphics[width=1.0\textwidth]{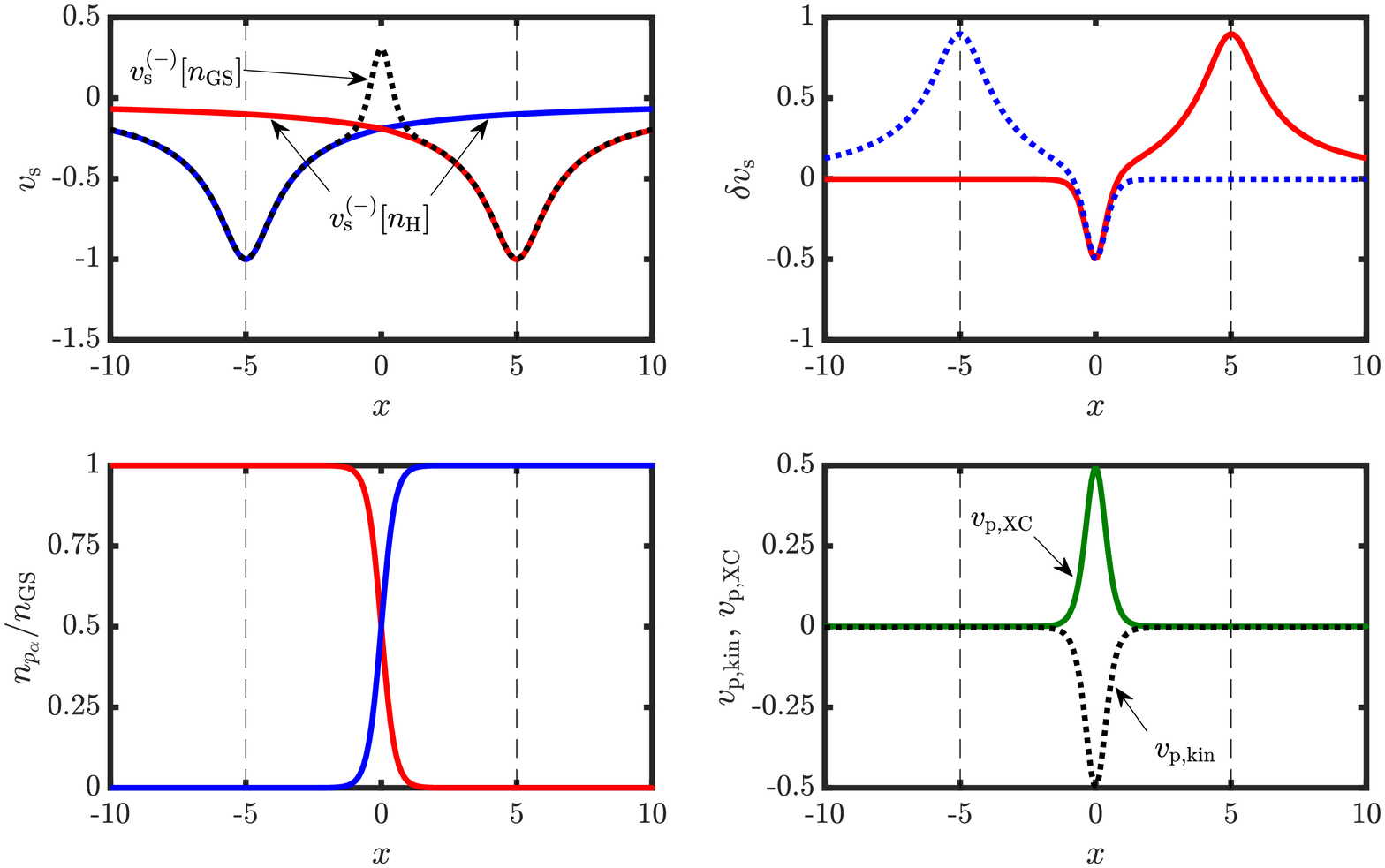}
\caption{The relationship between features of $v_{\mathrm{p},\mathrm{kin}}(x)$ and the peak of molecular $v_{\mathrm{s}}(x)$ for $\mathrm{H_2}$ model at $\lambda=1$ and $R=10.0$. Top left: molecular KS potential $v_\mathrm{s}^{(-)}[n_{\mathrm{GS}}](x)$ and fragment KS potentials $v_\mathrm{s}^{(-)}[n_{\mathrm{H}}](x)$. Top right: the differences between the molecular and fragment potentials. Bottom left: ${\npal(x)}/{n_{\mathrm{GS}}(x)}$ terms. Bottom right: kinetic and XC contributions to the partition potential. Vertical dashed lines indicate the position of nuclei.}
\label{fig:2} 
\end{figure*}

It may appear that the contributions from $v_{\mathrm{p},\mathrm{kin}}(x)$ and $v_{\mathrm{p},\mathrm{XC}}(x)$ at large separation are unimportant as they cancel each other. However, the bottom right panel in fig. \ref{fig:2} shows that these features have high magnitude. Since in practice $v_{\mathrm{p},\mathrm{kin}}(x)$ and $v_{\mathrm{p},\mathrm{XC}}(x)$ are approximated separately, the accuracy of the total $\vp(x)$ can be highly sensitive to the errors in these approximations.

In addition, fig. \ref{fig:2} shows the formation of $v_{\mathrm{p},\mathrm{kin}}(x)$ according to eq. \ref{subeq:vpdc2a}. Top left panel shows $v_\mathrm{s}^{(-)}[n_{\mathrm{GS}}](x)$ along with $v_\mathrm{s}^{(-)}[n_{\mathrm{H}}](x)$'s. We observe that $v_\mathrm{s}^{(-)}[n_{\mathrm{H}}](x)$ matches closely with $v_\mathrm{s}^{(-)}[n_{\mathrm{GS}}](x)$ in the nuclear regions. The difference between the fragment and molecular KS potentials $\delta v_{\mathrm{s}}(x)$, plotted at the top right, has the flat region around their nucleus. The differences are weighted by the corresponding ${\npal(x)}/{n_{\mathrm{GS}}(x)}$ terms and summed, producing the total $v_{\mathrm{p},\mathrm{kin}}(x)$. We note that $v_{\mathrm{p},\mathrm{kin}}(x)$ has a well from the peak in $v_\mathrm{s}^{(-)}[n_{\mathrm{GS}}](x)$. The peak in $v_{\mathrm{p},\mathrm{XC}}(x)$ has the \textit{same} origin \cite{GB96,HTR09,TMM09,FNRM16} and it nearly cancels the well in $v_{\mathrm{p},\mathrm{kin}}(x)$. This cancelation is not exact and the fine features in $v_{\mathrm{p},\mathrm{kin}}(x)+v_{\mathrm{p},\mathrm{XC}}(x)$ can still be observed.
\begin{figure*}
\includegraphics[width=1.0\textwidth]{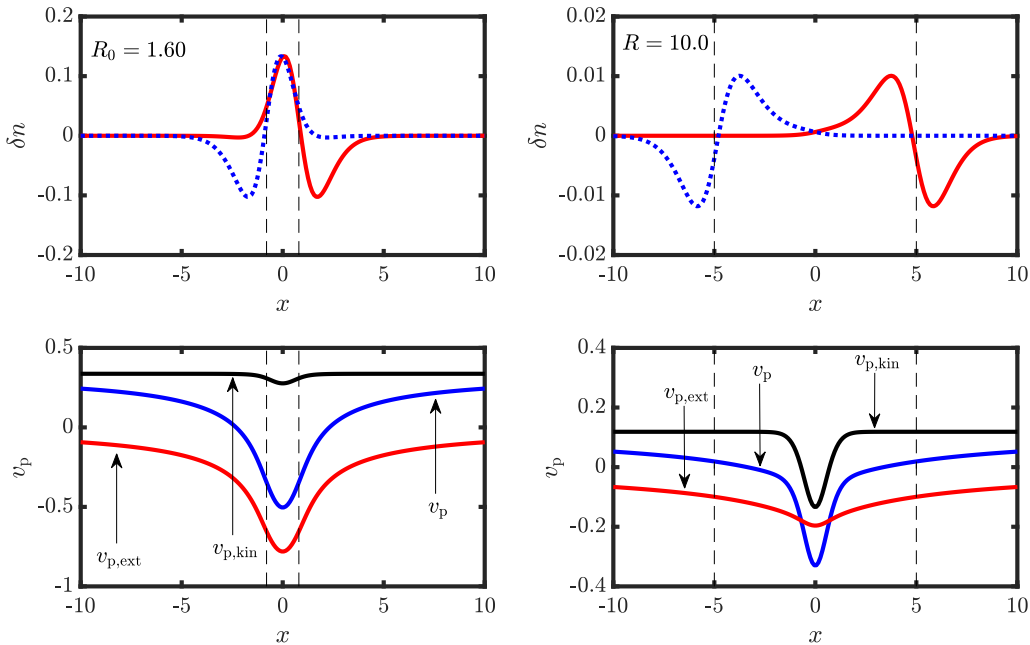}
\caption{Same as fig. \ref{fig:1}, but for $\lambda=0$}
\label{fig:3} 
\end{figure*}

We turn off the electron-electron interaction in the system by setting $\lambda=0$. The results are shown in fig. \ref{fig:3}. Our method recovers the trivial result that $v_\mathrm{p,H}(x)$ and $v_\mathrm{p,XC}(x)$ are zero. At both separations, $\vp(x)$ has a single well. At equilibrium, this well is dominated by $v_{\mathrm{p},\mathrm{ext}}(x)$. In contrast, at $R=10.0$, the well is predominantly determined by $v_{\mathrm{p},\mathrm{kin}}(x)$. The $\vp(x)$ plots are consistent with previously reported ones for noninteracting systems, \cite{CWCB09,TNW12} but the present work shows that the well in $\vp(x)$ is dominated by different components at \textit{different} internuclear distances.

\paragraph{$\mathbf{HeH^+}$ model:} We study the features of $\vp(x)$ in the simplest heteronuclear molecular ion $\mathrm{HeH^+}$ at equilibrium separation. This model has non-integer optimal occupations. We use this fact to analyze the relationship between the kinetic component of $\vp(x)$ and the KS gap of PT fragments.
\begin{figure*}
\includegraphics[width=1.0\textwidth]{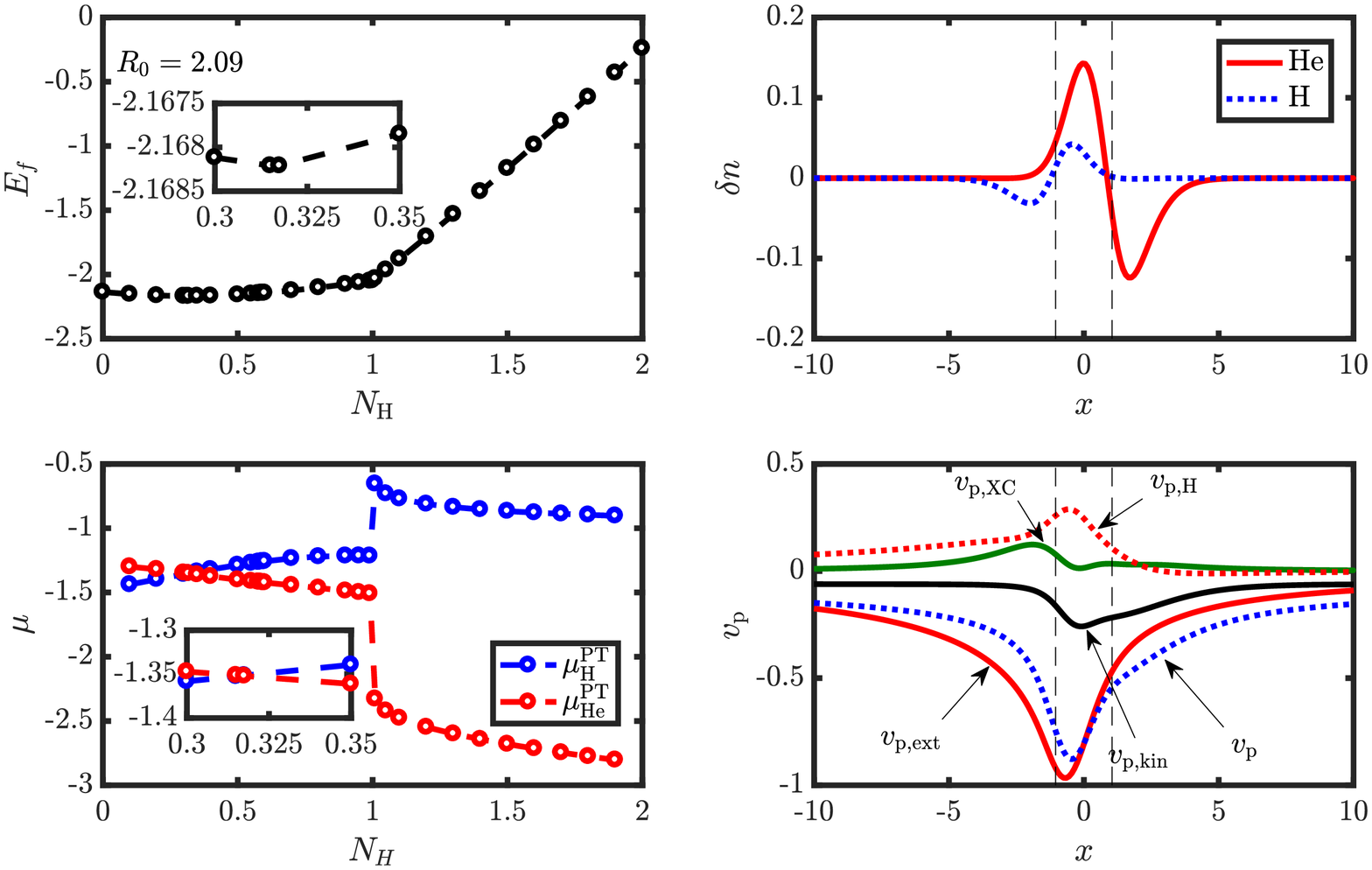}
\caption{Summary of the PT results for the model system of $\mathrm{HeH^+}$ at equilibrium separation and $\lambda=1$. Left: fragment energies (top) and PT chemical potentials (bottom) at varying occupations on $\mathrm{H}$ atom. Right: density deformations relative relative to the isolated fragments with the optimal electron occupations (top) and corresponding partition potential along its components (bottom). Vertical dashed lines indicate the position of nuclei ($\mathrm{H}$ is on the left).}
\label{fig:4} 
\end{figure*}

The left two panels of fig. \ref{fig:4} show the behavior of $E_{\mathrm{f}}[\sna]$ as a function of the number of electrons on the hydrogen atom, at the equilibrium bond distance of $2.09$ $a.u.$ The curvature of the energy plot is an important consequence of accounting for the finite-distance interfragment interactions (in contrast, the plot of energy versus the number of electrons in DFT consists of straight-line segments). This curvature does not smoothen the graph at integer occupations, where it still has a cusp. The graph has a minimum when $N_{\mathrm H} \approx 0.3175$. At this occupation, we also observe the chemical potential equalization of the fragments. A rigorous definition of fragments allows the discussion of the nature of a chemical bond and the optimal occupations suggest the amount of the ionic character a bond has. The connection between 1-D models and real bonds is, of course, not obvious. More generally, the physical interpretation of $\mathrm{PT}$ fragment properties is still an open question.

The top right panel of fig. \ref{fig:4} shows the density deformations relative to the isolated fragments with the optimal electron occupations. We observe that both $\mathrm{He}$ and $\mathrm{H}$ densities are shifted towards the interatomic region. The partition potential that facilitates this shift is plotted at the bottom right of fig. \ref{fig:4}, along with its components. Although its overall shape is similar to $\mathrm{H_2}$ at equilibrium bond distance, $\vp(x)$ of $\mathrm{HeH^+}$ is dominated by $v_{\mathrm{p},\mathrm{ext}}(x)$. Naively, this can be attributed to the fact that $\mathrm{HeH^+}$ is an ion and the electron-nuclear interactions are the dominant ones.

The non-integer occupation numbers allow to establish the relationship between $v_{\mathrm{p},\mathrm{kin}}(x)$ and the fragment KS gaps $\mathrm{\Delta}^{\alpha}=\mathrm{I}^{\alpha}-\mathrm{A}^{\alpha}$, where $\mathrm{I}^{\alpha}$ is the ionization potential and $\mathrm{A}^{\alpha}$ is the electron affinity of a fragment in the presence of $\vp(x)$. If we assume the near-linearity of the fragment KS potentials, \cite{GT14} eq. \ref{subeq:vpdc2a} can be approximated as $v_{\mathrm{p},\mathrm{kin}}(x)\approx v_{\mathrm{p},\mathrm{kin}}^{\mathrm{nl}}(x)$, where:
\begin{equation}
    \label{eq:vpkinapp}
    \begin{split}
    v_{\mathrm{p},\mathrm{kin}}^{\mathrm{nl}}(x)=&\sum_\sa\biggl\{ \frac{n_\sa(x)}{n_{\mathrm{GS}}(x)}v_\mathrm{s}[n_\sa](x)-\\
    &(1-\wa)\mathrm{\Delta}^{\alpha}\mathcal{Q}_{p_\sa}(x,x) \biggr\}-v_\mathrm{s}^{(-)}[n_{\mathrm{GS}}](x).
    \end{split}
\end{equation}
\begin{figure*}
\includegraphics[width=1.0\textwidth]{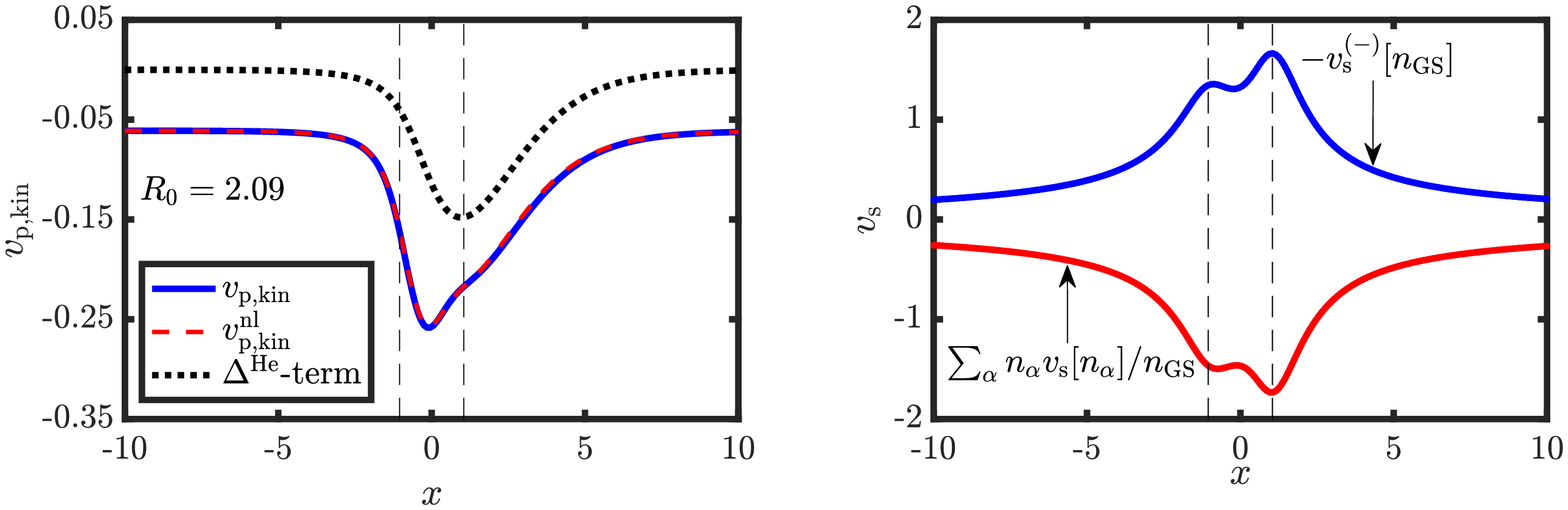}
\caption{The relationship between $v_{\mathrm{p},\mathrm{kin}}$ and $\mathrm{\Delta}^{\mathrm{He}}$ defined through eq. \ref{eq:vpkinapp}. $\mathrm{\Delta}^{\mathrm{He}}$-term stands for $-(1-\omega_\mathrm{He})\mathrm{\Delta}^{\mathrm{He}}\mathcal{Q}_{p_\mathrm{He}}(x,x)$. Vertical dashed lines indicate the position of nuclei ($\mathrm{H}$ is on the left).}
\label{fig:5} 
\end{figure*}
Fig. \ref{fig:5} indicates that this approximation is in excellent agreement with the exact $ v_{\mathrm{p},\mathrm{kin}}(x)$. The right panel in fig. \ref{fig:5} compares the molecular KS potential to the weighted sum of the fragment KS potentials, $\sum_\alpha{n_\alpha [n_\alpha](x)}/{n_{\mathrm{GS}}(x)}v_\mathrm{s}(x)$ from eq. \ref{eq:vpkinapp}. We can see that these two contributions almost entirely cancel out and $v_{\mathrm{p},\mathrm{kin}}(x)$ is largely determined by the $(1-\omega_\mathrm{He})\mathrm{\Delta}^{\mathrm{He}}\mathcal{Q}_{p_\mathrm{He}}(x,x)$ term (note that there is no contribution from $\mathrm{\Delta}^{\mathrm{H}}$ because $p_\mathrm{H}=0$). Additional calculations on model systems suggest that the fragment KS term closely mimics $-v_\mathrm{s}^{(-)}[n_{\mathrm{GS}}](x)$ in the high density regions, but it misses its low density peak-and-step features.

\paragraph{$\mathbf{LiH}$ model:} We consider a heteroatomic dimer model of lithium hydride that separates into neutral fragments. In this model, the core electrons are not treated explicitly but their effects are simulated by adjusting the parameters of the external potential function. The modified electronic Hamiltonian of eq. \ref{eq:H} is:
\begin{equation}
\begin{split}
\mathcal{H} =& \sum_{i=1,2}\biggl\{-\frac{1}{2}\nabla^2_{x_i} - \frac{1}{\sqrt{2.25+(x_i-R_{\mathrm{Li}})^2}} - \\ &\frac{Z_{\mathrm{X}}}{\sqrt{0.6+(x_i-R_\mathrm{H})^2}}\biggr\}+\frac{1}{\sqrt{0.7+(x_1-x_2)^2}},
\end{split}
\label{eq:HLiH}
\end{equation}
where the SC parameters for $\mathrm{Li}$, $\mathrm{H}$ and electron-electron interactions (2.25, 0.70 and 0.60 respectively) are chosen following the same considerations as in ref. \cite{TMM09}. These parameters produce the correct ionization potential difference between isolated $\mathrm{Li}$ and $\mathrm{H}$ atoms. The individual ionization potentials produced by this model are higher than the real ones, making the densities less diffuse and allowing us to use a simulation box of $25$ $\mathrm{a.u}$. 
\begin{figure*}
\includegraphics[width=1.0\textwidth]{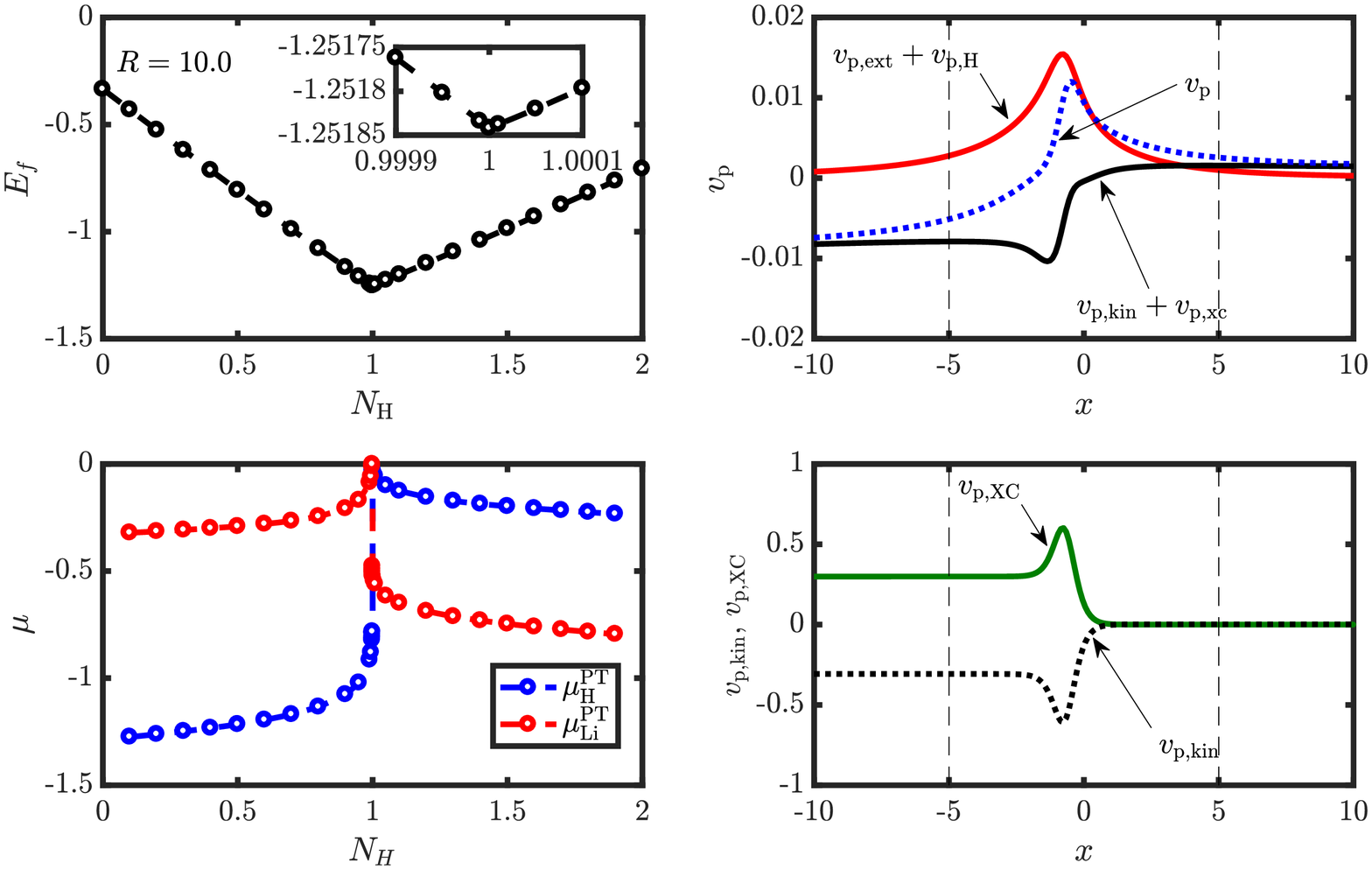}
\caption{Summary of the PT results for the model system of $\mathrm{LiH}$, defined through eq. \ref{eq:HLiH} at $R=10.0$. Left: fragment energies (top) and PT chemical potentials (bottom) at varying occupations on $\mathrm{H}$ atom. Right: partition potential and its components (top); kinetic and XC contributions to $\vp(x)$ (bottom). Vertical dashed lines indicate the position of nuclei ($\mathrm{H}$ is on the left)}
\label{fig:6} 
\end{figure*}

The results for $\mathrm{LiH}$ are summarized in fig. \ref{fig:6}. The left two graphs show the fragment energies and chemical potentials at varying occupation numbers. $E_{\mathrm{f}}$ is minimized when $N_{\mathrm{H}}$ (and obviously $N_{\mathrm{Li}}$) is equal to 1. This point is a cusp in $E_{\mathrm{f}}$ as expected from eq. \ref{eq:Ef}. $R= 10.0$ $a.u.$ can be taken as the large separation limit in our model and it shows that the bond breaking is homolytic. Although not obvious from the plot, the graph of $E_{\mathrm{f}}$ is curved, similar to the one for $\mathrm{HeH^+}$ in fig. \ref{fig:4}. The chemical potentials exhibit a step-like feature into integer occupations, which prevent the condition of eq. \ref{eq:mu} to be satisfied. The right two graphs show $\vp(x)$ and its decomposition. Similarly to the case of $\mathrm{H_2}$, $\vp(x)$ has a peak in the internuclear region, dominated by the $v_{\mathrm{p},\mathrm{ext}}(x)+v_{\mathrm{p},\mathrm{H}}(x)$ term. Moreover, the $v_{\mathrm{p},\mathrm{kin}}(x)$ and $v_{\mathrm{p},\mathrm{XC}}(x)$ almost completely cancel out. Analogously to the case of $\mathrm{H_2}$, their features are connected to the features of the molecular KS potential. \cite{GB96,HTR09,TMM09,FNRM16} In addition to the peak, in this case,  $v_{\mathrm{p},\mathrm{kin}}(x)$ and $v_{\mathrm{p},\mathrm{XC}}(x)$ also display a step. The steps almost entirely cancel out. The remaining small peak we observe in the top right panel of fig. \ref{fig:6} is likely due to the long range nature of SC potentials.
%%%%%%%%
%Conclusion %
%%%%%%%%
\section{Concluding remarks}
In spite of the simplicity of this model, we expect the same features discovered here to be present in real molecules. Explicit treatment of core electrons and 3D-Coulomb interactions would be of course needed to verify this.

Finally, the decomposition of $\vp(x)$ through eq. \ref{eq:vpdc} provides a useful way for identifying the origin of important features of $\vp(x)$ and linking them to the approximations used in practical density-embedding calculations. We plan to investigate in future work the extent to which approximate XC and non-additive kinetic energy functionals reproduce the features of $\vp(x)$ observed here.
%%%%%%%%%%%%
%Acknowledgements %
%%%%%%%%%%%%
\section{Acknowledgements}
We thank Hardy Gross for asking the question that inspired this work. We still owe him pictures of the exact $\vp(r)$ for a \textit{real} molecule. We acknowledge support from the National Science Foundation CAREER program under Grant No. CHE-1149968.
\bibliographystyle{spphys}
\bibliography{bibliographyPaper}   % name your BibTeX data base

\end{document}